\documentstyle[epsf]{l-aa}

\begin{document}

\thesaurus{10 (10.11.1; 8.11.1; 5.18.1)}

\title{Derivation of the Galactic rotation curve
       using space velocities}

\author{S.\ Frink \and B.\ Fuchs \and S.\ R\"oser \and R.\ Wielen}
\offprints{S.\ Frink}
\institute{Astronomisches Rechen-Institut Heidelberg, M\"onchhofstra{\ss}e
 12-14, D-69120 Heidelberg, Germany}

\date{Received ; accepted }

\maketitle

\begin{abstract}
We present rotation curves of the Galaxy based on the space-velocities of
197 OB stars and 144 classical cepheids, respectively, which range over a
galactocentric distance interval of about 6 to 12\,kpc. No significant
differences between these rotation curves and rotation curves based solely
on radial velocities assuming circular rotation are found. We derive an
angular velocity of the LSR of $\Omega_0 = 5.5 \pm 0.4$\,mas/a (OB stars)
and $\Omega_0 = 5.4 \pm 0.5$\,mas/a (cepheids), which is in
agreement with the IAU 1985 value of $\Omega_0 = 5.5$\,mas/a.
If we correct for probable rotations of the FK5 system, the corresponding
angular velocities are $\Omega_0 = 6.0$\,mas/a (OB stars) and $\Omega_0 =
6.2$\,mas/a (cepheids). These values agree better with the value of
$\Omega_0 = 6.4$\,mas/a derived from the VLA measurement of the proper
motion of Sgr\,A$^{*}$.
\keywords{Galaxy: kinematics and dynamics -- Stars: kinematics -- Reference
systems}
\end{abstract}

\begin{figure}
\begin{center}
\epsfxsize=7cm
\leavevmode
\epsffile{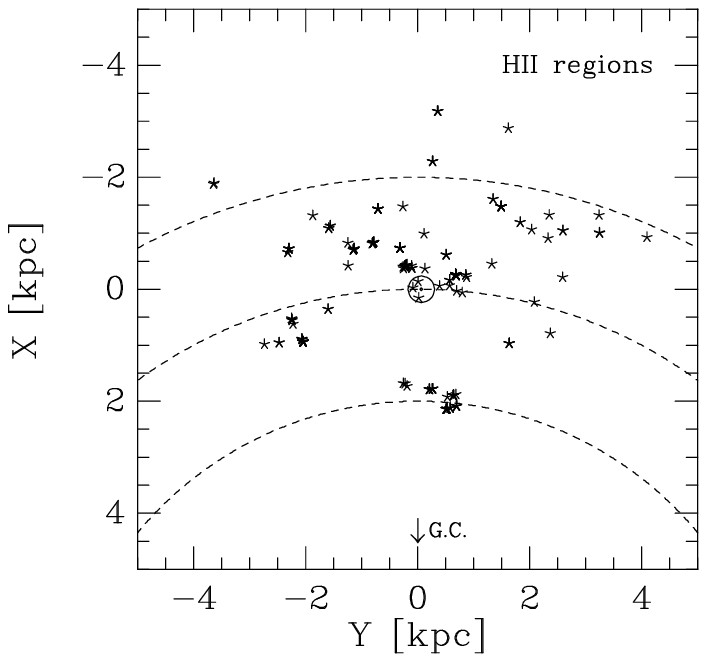}
\end{center}
\begin{center}
\epsfxsize=7cm
\leavevmode
\epsffile{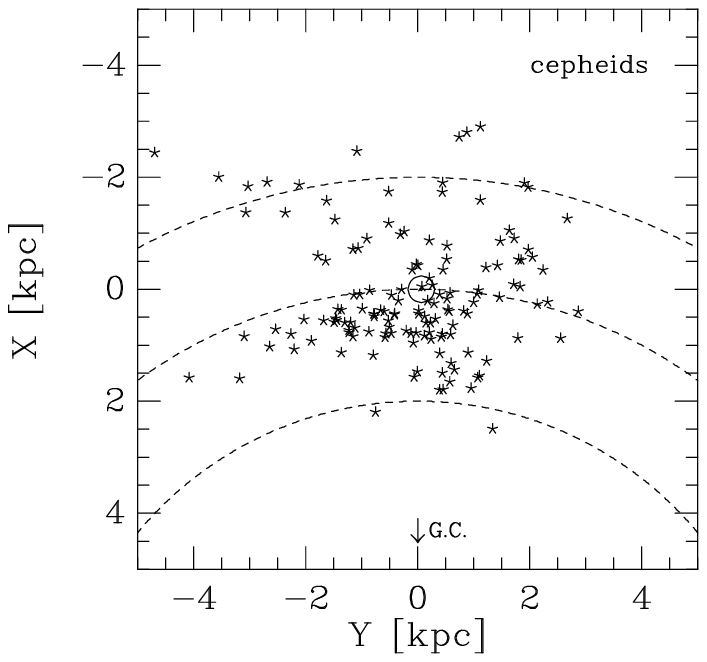}
\end{center}
\caption[]{\label{xyz} Positions of the H\,{\sc ii} regions (upper panel) and
the cepheids (lower panel), for which the proper motions are well-known,
projected onto the galactic plane. The position of the Sun is in the centre
of the panels. The dashed lines indicate circles
around the galactic centre with radii 6.5, 8.5 and 10.5 kpc, respectively.}
\end{figure}

\begin{figure*}
\begin{center}
\leavevmode
\epsffile{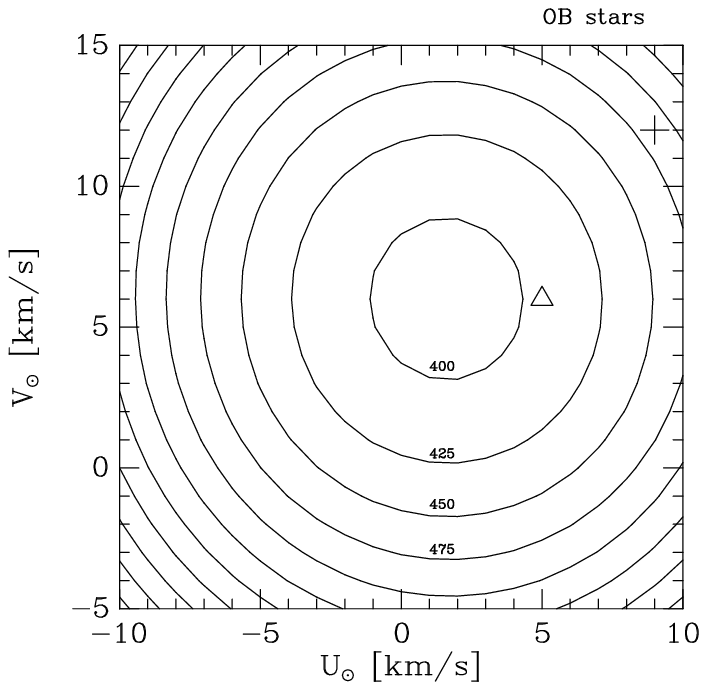}
\hspace*{3em}
\epsffile{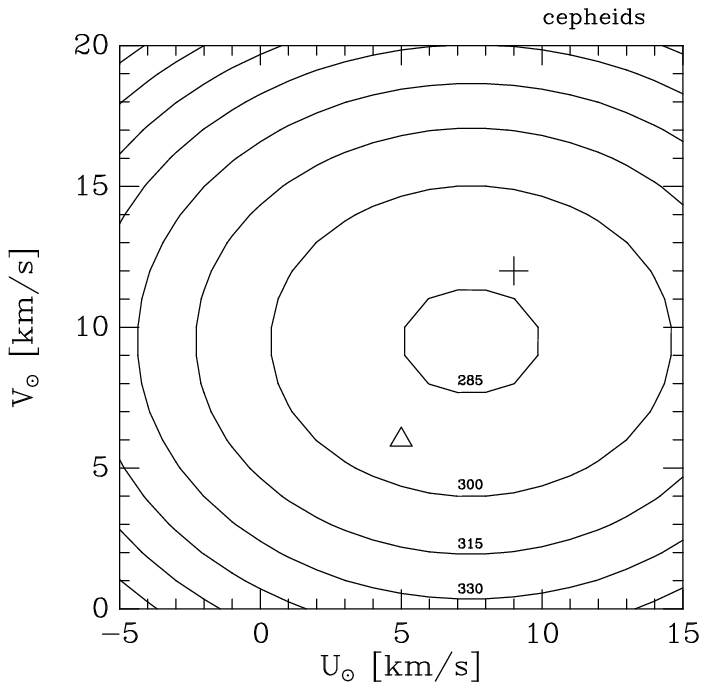}
\end{center}
\caption[]{\label{grid} Contourmaps of the numerical value of $\chi^2$ as a
function of the solar motion $U_0$ and $V_0$. A cross indicates the standard
value of $U_0 = 9$\,km/s, $V_0 = 12$\,km/s, a triangle the value
$U_0 = 5$\,km/s, $V_0 = 6$\,km/s used in the present study. Note that the
contours are steeper in the fit to the OB star data (left panel) than
in the fit to the cepheid data (right panel).}
\end{figure*}

\section{Introduction}
The galactic rotation curve has been determined for the inner parts of the
galactic disk, interior to the solar annulus, from H\,{\sc i}-measurements
using the tangential point method (Burton \& Gordon 1978), whereas the outer
rotation curve has been determined using radial velocities of objects with
individually known distances, e.\,g.\ OB stars (Fich et al.\ 1989),
planetary nebulae (Schneider \& Terzian 1983),
young open clusters (Hron 1987) and carbon stars (Metzger \&
Schechter 1994). An alternative method is based on the vertical
thickness of the galactic H\,{\sc i}-layer (Merrifield 1992). Recently
Brand \& Blitz (1993) have rederived the outer rotation curve from CO radial
velocities of OB stars associated with H\,{\sc ii} regions, and Pont
et al.\ (1994) have used new radial velocity measurements of classical
cepheids for this purpose. All these methods rely on the assumption of circular
orbits around the galactic centre, so that radial velocities can be converted
to circular velocities.
Obviously proper motions of the objects may provide independent information
on the rotation curve. The PPM catalogue which has recently been compiled
at the Astronomisches Rechen-Institut (R\"oser \& Bastian 1991, Bastian et
al.\ 1993, R\"oser et al.\ 1994) is well suited as a broad data
base of proper motions of high accuracy for such purposes.

\section{Data}
\subsection{OB stars}
Brand \& Blitz (1993) give a list containing radial velocities of
H\,{\sc ii} regions and their associated
molecular cloud complexes which is the basis of our analysis. The exciting
OB stars in the northern hemisphere are tabulated by Georgelin (1975) and
others (see Blitz et al.\ 1982 for references) and
can be identified by cross reference numbers. For the exciting OB stars of
the H\,{\sc ii} regions in the southern hemisphere finding charts are given
by Brand (1986). Using cross reference numbers, spectral types,
magnitudes or accurate stellar positions, which we derived in an intermediate
identification step using the HST Guide Star Catalog (Lasker et al.\ 1990),
we have identified as many stars from the source lists in the PPM
Star Catalogue as possible.

In total we have been able to identify 228 OB stars associated
with 71 H\,{\sc ii} regions.

\subsection{Cepheids}
Radial velocities and distances of classical cepheids are taken from the
recent work of Pont et al.\ (1994). Using again accurate stellar positions
and cross reference numbers, which we obtained from the HIPPARCOS Input
Catalogue (Turon et al.\ 1992) and from Kholopov et al.\ (1985-1987), we
were able to identify 152 cepheids in the PPM Star Catalogue.

\smallskip
The PPM Star Catalogue is tied into the FK5 system (Fricke et al.\ 1988).
Lindegren et al.\ (1995) have shown by a systematic comparison of the
FK5 system with preliminary HIPPARCOS data that the FK5 positional system
has still deficiencies, in particular in a strip at about
$\delta\approx-40^{\circ}$, corresponding to $l\approx260^{\circ},\;
b\approx0^{\circ}$.
Unfortunately a number of H\,{\sc ii} regions and cepheids fall into this
area, which we have omitted from our analysis because of this systematic
uncertainty, leaving us with 197 OB stars associated with 59 H\,{\sc ii}
regions and 144 cepheids with well-determined proper motions for our final
analysis. Tables \ref{obtab} and \ref{ceptab} give the identification numbers
of these stars in the PPM catalogue.

Using the positions and proper motions of the stars given in the PPM Star
Catalogue and the distances and the radial velocities given by Brand \&
Blitz (1993) or Pont et al.\ (1994), respectively, we have
calculated the positions and the velocities of the stars in galactic
coordinates. Fig.\,\ref{xyz} shows the projection of the positions of
the objects onto the galactic plane. $X$ and $Y$ are the cartesian spatial
coordinates of the stars relative to the Sun. The $X$-axis points towards
the galactic centre and the $Y$-axis into the direction of galactic
rotation.

\begin{table*}
\caption[]{\label{obtab}
 H\,{\sc ii} regions and their associated OB stars.
 The first column gives the Sharpless or Brand identification numbers of
 the H\,{\sc ii} regions. The second column gives the identification
 numbers of the OB stars in this region which could be found in the
 PPM Star Catalogue.}
\begin{flushleft}
\begin{center}
\begin{minipage}[b]{4.3cm}
\begin{tabular}{@{$\quad$}lr}
\noalign{\smallskip}
\hline
\noalign{\smallskip}
\multicolumn{1}{c}{H\,{\sc ii}} & \multicolumn{1}{c}{PPM} \\
\noalign{\smallskip}
\hline
\noalign{\smallskip}
S8       & 296225 \\
         & 296206 \\
S11      & 296337 \\
S25      & 267779 \\
         & 267782 \\
         & 267822 \\
         & 267797 \\
S29      & 267950 \\
S31      & 267973 \\
S32      & 267965 \\
         & 267970 \\
         & 267964 \\
S27      & 231915 \\
S41      & 234340 \\
         & 234237 \\
         & 234243 \\
         & 234273 \\
         & 234280 \\
         & 234293 \\
         & 234337 \\
         & 234362 \\
S44      & 234275 \\
S45      & 234399 \\
         & 234410 \\
S46      & 234023 \\
S49      & 234332 \\
         & 234333 \\
         & 234345 \\
         & 234348 \\
S54      & 234320 \\
         & 234299 \\
         & 234306 \\
         & 234321 \\
         & 234336 \\
         & 234352 \\
         & 234357 \\
S86      & 109578 \\
         & 109560 \\
         & 109575 \\
S101     &  83984 \\
S112     &  60147 \\
S117     &  60726 \\
S119     &  61260 \\
S124     &  39725 \\
S126     &  88031 \\
S129     &  39225 \\
S132     &  40485 \\
S134     &  40324 \\
S137     &  23309 \\
         &  23333 \\
\noalign{\smallskip}
\hline
\end{tabular}
\end{minipage}
\begin{minipage}[b]{4.3cm}
\begin{tabular}{@{$\quad$}lr}
\noalign{\smallskip}
\hline
\noalign{\smallskip}
\multicolumn{1}{c}{H\,{\sc ii}} & \multicolumn{1}{c}{PPM} \\
\noalign{\smallskip}
\hline
\noalign{\smallskip}
S140     &  23638 \\
         &  23640 \\
S142     &  41076 \\
         &  41098 \\
S154     &  23983 \\
S155     &  24016 \\
         &  24060 \\
         &  24070 \\
         &  24074 \\
         &  24030 \\
         &  24072 \\
         &  23979 \\
         &  24019 \\
         &  24021 \\
         &  24036 \\
         &  24043 \\
         &  24049 \\
         &  24098 \\
         &  24093 \\
         &  24117 \\
         &  24148 \\
         &  24150 \\
         &  24162 \\
         &  24169 \\
         &  24185 \\
         &  24186 \\
S157     &  24306 \\
S161     &  24309 \\
         &  24329 \\
S162     &  24384 \\
S170     &  11863 \\
S173     &  12143 \\
S184     &  25788 \\
         &  25791 \\
S190     &  13731 \\
         &  13702 \\
         &  13720 \\
         &  13721 \\
         &  13723 \\
         &  13712 \\
         &  13718 \\
S199     &  28316 \\
         &  28326 \\
S202     &  14174 \\
         &  14185 \\
         &  14227 \\
S206     &  28874 \\
S220     &  68961 \\
S232     &  70744 \\
S234     &  70403 \\
\noalign{\smallskip}
\hline
\end{tabular}
\end{minipage}
\begin{minipage}[b]{4.3cm}
\begin{tabular}{@{$\quad$}lr}
\noalign{\smallskip}
\hline
\noalign{\smallskip}
\multicolumn{1}{c}{H\,{\sc ii}} & \multicolumn{1}{c}{PPM} \\
\noalign{\smallskip}
\hline
\noalign{\smallskip}
S234     &  70368 \\
         &  70374 \\
S236     &  70266 \\
         &  70271 \\
         &  70273 \\
S238     & 119824 \\
S252     &  95519 \\
S263     & 148818 \\
S264     & 149155 \\
         & 149166 \\
S273     & 151033 \\
         & 151058 \\
         & 151028 \\
         & 151013 \\
         & 151053 \\
         & 151030 \\
         & 151073 \\
         & 151050 \\
S275     & 150705 \\
         & 150694 \\
         & 150692 \\
         & 150670 \\
         & 150682 \\
         & 150706 \\
S277     & 188303 \\
S279     & 188224 \\
S281     & 187839 \\
         & 175945 \\
         & 188455 \\
         & 175888 \\
         & 188218 \\
         & 188223 \\
         & 188225 \\
         & 188149 \\
         & 215598 \\
S292     & 218092 \\
S293     & 218026 \\
S295     & 218051 \\
S296     & 218372 \\
         & 218096 \\
         & 218138 \\
         & 218164 \\
         & 218171 \\
         & 218242 \\
         & 218262 \\
         & 218324 \\
S297     & 218121 \\
S310     & 252114 \\
S311     & 253358 \\
         & 253404 \\
\noalign{\smallskip}
\hline
\end{tabular}
\end{minipage}
\begin{minipage}[b]{4.3cm}
\begin{tabular}{@{$\quad$}lr}
\noalign{\smallskip}
\hline
\noalign{\smallskip}
\multicolumn{1}{l}{\enspace H\,{\sc ii}} & \multicolumn{1}{c}{PPM} \\
\noalign{\smallskip}
\hline
\noalign{\smallskip}
BBW16    & 727432 \\
         & 252138 \\
         & 727439 \\
         & 727444 \\
BBW17B   & 252139 \\
         & 252148 \\
         & 252158 \\
BBW23    & 252308 \\
         & 252289 \\
         & 252363 \\
         & 252304 \\
BBW29    & 252486 \\
BBW104B  & 284824 \\
BBW106   & 740844 \\
BBW127   & 285005 \\
         & 740995 \\
         & 740981 \\
         & 740980 \\
BBW133   & 285078 \\
         & 741033 \\
BBW283   & 769065 \\
         & 338612 \\
BBW300B  & 338995 \\
         & 339034 \\
         & 339041 \\
         & 338930 \\
BBW316D  & 339247 \\
BBW347   & 339851 \\
BBW348A  & 358562 \\
         & 358560 \\
BBW362A  & 358835 \\
         & 358846 \\
         & 358849 \\
         & 358855 \\
         & 358857 \\
         & 358862 \\
         & 358863 \\
         & 358865 \\
         & 358869 \\
         & 358872 \\
         & 358875 \\
         & 358876 \\
         & 358883 \\
         & 358884 \\
BBW362C  & 358747 \\
         & 358742 \\
BBW362F  & 358755 \\
\noalign{\smallskip}
\hline
         &        \\
         &        \\
         &        \\
\end{tabular}
\vfill
\end{minipage}
\end{center}
\end{flushleft}
\end{table*}
\vfill

\begin{table*}
\caption[]{\label{ceptab}
 Cepheids used to calculate the rotation curve.}
\begin{flushleft}
\begin{center}
\begin{minipage}[b]{4.3cm}
\begin{tabular}{@{\quad}r@{\,}lr}
\noalign{\smallskip}
\hline
\noalign{\smallskip}
\multicolumn{2}{c}{cepheid} & \multicolumn{1}{c}{PPM} \\
\noalign{\smallskip}
\hline
\noalign{\smallskip}
   $\eta $ & Aql  & 168843  \\
         U & Aql  & 202954  \\
        SZ & Aql  & 167110  \\
        TT & Aql  & 167242  \\
        FF & Aql  & 135550  \\
        FM & Aql  & 135861  \\
        FN & Aql  & 167384  \\
      V336 & Aql  & 167002  \\
      V496 & Aql  & 202574  \\
      V600 & Aql  & 167670  \\
         Y & Aur  &  48141  \\
        RT & Aur  &  71665  \\
        RX & Aur  &  69838  \\
        SY & Aur  &  47840  \\
        RY & CMa  & 218422  \\
        RZ & CMa  & 713898  \\
        SS & CMa  & 252386  \\
        TV & CMa  & 713601  \\
        TW & CMa  & 713915  \\
        VZ & CMa  & 727576  \\
        RW & Cam  &  28771  \\
        RX & Cam  &  28898  \\
   $\iota$ & Car  & 357533  \\
         U & Car  & 339614  \\
         V & Car  & 356744  \\
         Y & Car  & 339156  \\
        UW & Car  & 339046  \\
        UX & Car  & 339090  \\
        XX & Car  & 358358  \\
        XY & Car  & 358418  \\
        XZ & Car  & 358450  \\
        YZ & Car  & 339070  \\
        ER & Car  & 339816  \\
        GI & Car  & 339882  \\
        IT & Car  & 358563  \\
        RY & Cas  &  42304  \\
\noalign{\smallskip}
\hline
\end{tabular}
\end{minipage}
\begin{minipage}[b]{4.3cm}
\begin{tabular}{@{\quad}r@{\,}lr}
\noalign{\smallskip}
\hline
\noalign{\smallskip}
\multicolumn{2}{c}{cepheid} & \multicolumn{1}{c}{PPM} \\
\noalign{\smallskip}
\hline
\noalign{\smallskip}
        SU & Cas  &  13918  \\
        SW & Cas  &  41496  \\
        SZ & Cas  &  27633  \\
        XY & Cas  &  12507  \\
        DL & Cas  &  12258  \\
        FM & Cas  &  25159  \\
      V636 & Cas  &  13051  \\
         V & Cen  & 342919  \\
        UZ & Cen  & 358897  \\
        XX & Cen  & 342137  \\
        BB & Cen  & 359040  \\
        BK & Cen  & 358996  \\
      V378 & Cen  & 360023  \\
      V381 & Cen  & 342310  \\
      V419 & Cen  & 340152  \\
 $\delta $ & Cep  &  40731  \\
        CR & Cep  &  41067  \\
        AX & Cir  & 361050  \\
         R & Cru  & 359375  \\
         S & Cru  & 341409  \\
         T & Cru  & 359350  \\
         X & Cru  & 341282  \\
        VW & Cru  & 359483  \\
        AG & Cru  & 341197  \\
        BG & Cru  & 341066  \\
         X & Cyg  &  85419  \\
        SU & Cyg  & 109630  \\
        SZ & Cyg  &  60114  \\
        TX & Cyg  &  60815  \\
        VX & Cyg  &  60748  \\
        VY & Cyg  &  85963  \\
        VZ & Cyg  &  62131  \\
        CD & Cyg  &  84139  \\
        DT & Cyg  &  86036  \\
      V386 & Cyg  &  61152  \\
      V532 & Cyg  &  61309  \\
\noalign{\smallskip}
\hline
\end{tabular}
\end{minipage}
\begin{minipage}[b]{4.3cm}
\begin{tabular}{@{\quad}r@{\,}lr}
\noalign{\smallskip}
\hline
\noalign{\smallskip}
\multicolumn{2}{c}{cepheid} & \multicolumn{1}{c}{PPM} \\
\noalign{\smallskip}
\hline
\noalign{\smallskip}
     V1334 & Cyg  &  86379  \\
  $\beta $ & Dor  & 354837  \\
  $\zeta $ & Gem  &  96982  \\
         W & Gem  & 122711  \\
         V & Lac  &  41123  \\
         X & Lac  &  41132  \\
         Y & Lac  &  40284  \\
         Z & Lac  &  40952  \\
        RR & Lac  &  40959  \\
        BG & Lac  &  62336  \\
        GH & Lup  & 343690  \\
         T & Mon  & 150465  \\
        SV & Mon  & 150342  \\
         R & Mus  & 359566  \\
         S & Mus  & 371417  \\
        RT & Mus  & 358941  \\
        UU & Mus  & 359025  \\
         S & Nor  & 344832  \\
         U & Nor  & 344085  \\
         Y & Oph  & 201153  \\
        BF & Oph  & 266398  \\
        RS & Ori  & 122354  \\
        SV & Per  &  47393  \\
        SX & Per  &  46949  \\
        VX & Per  &  27145  \\
        AW & Per  &  69651  \\
      V440 & Per  &  27565  \\
         X & Pup  & 252643  \\
        RS & Pup  & 284941  \\
        VX & Pup  & 252632  \\
        VZ & Pup  & 727776  \\
        WX & Pup  & 252971  \\
        WZ & Pup  & 728158  \\
        AQ & Pup  & 253563  \\
        AT & Pup  & 284926  \\
         S & Sge  & 137241  \\
\noalign{\smallskip}
\hline
\end{tabular}
\end{minipage}
\begin{minipage}[b]{4.3cm}
\begin{tabular}{@{\quad}r@{\,}lr}
\noalign{\smallskip}
\hline
\noalign{\smallskip}
\multicolumn{2}{c}{cepheid} & \multicolumn{1}{c}{PPM} \\
\noalign{\smallskip}
\hline
\noalign{\smallskip}
         U & Sgr  & 234663  \\
         W & Sgr  & 267817  \\
         X & Sgr  & 267303  \\
         Y & Sgr  & 234421  \\
        WZ & Sgr  & 234281  \\
        XX & Sgr  & 234503  \\
        YZ & Sgr  & 235075  \\
        AP & Sgr  & 268062  \\
        AY & Sgr  & 719227  \\
        BB & Sgr  & 268971  \\
      V350 & Sgr  & 268853  \\
        RV & Sco  & 295723  \\
        RY & Sco  & 296879  \\
        KQ & Sco  & 762230  \\
      V482 & Sco  & 296427  \\
      V500 & Sco  & 296821  \\
      V636 & Sco  & 322879  \\
         X & Sct  & 234646  \\
         Y & Sct  & 201965  \\
         Z & Sct  & 202059  \\
        RU & Sct  & 202038  \\
        SS & Sct  & 202078  \\
        EV & Sct  & 707164  \\
        ST & Tau  & 121352  \\
        SZ & Tau  & 120081  \\
         R & Tra  & 361350  \\
         S & Tra  & 361800  \\
         U & Tra  & 361877  \\
 $\alpha $ & UMi  &    431  \\
         V & Vel  & 337875  \\
        SV & Vel  & 339405  \\
        BG & Vel  & 337630  \\
         T & Vul  & 112020  \\
         U & Vul  & 109337  \\
         X & Vul  & 110094  \\
        SV & Vul  & 109871  \\
\noalign{\smallskip}
\hline
\end{tabular}
\end{minipage}
\end{center}
\end{flushleft}
\vfill
\end{table*}

\begin{table*}
\caption[]{\label{precession} Effects of the correction of the constant of
precession by $\Delta p_1 = -3.2$\,mas/a and different corrections of the
motion of the equinox $\Delta e$ on the angular velocities $\Omega_0$,
$\Omega_1$ and $\Omega_2$. $\Omega_0$ is the angular velocity of the LSR and
$\sqrt{\Omega_1^2 + \Omega_2^2}$ is the tilting of the galactic plane which
should be minimized by adjusting $\Delta e$.
Only OB stars with distances greater than 800\,pc from
the sun are considered in order to avoid contamination by Gould's Belt.
The mean error of an angular velocity component is about 0.5\,mas/a.}
\begin{flushleft}
\begin{center}
\begin{minipage}[b]{8.7cm}
\begin{center}  OB stars \hphantom{cepheids} \end{center}
\vspace*{-2ex}
\begin{tabular}{ccccc}
\noalign{\smallskip}
\hline
\noalign{\smallskip}
\multicolumn{1}{c}{$\Delta e$} & \multicolumn{1}{c}{$\Omega_0$} &
\multicolumn{1}{c}{$\Omega_1$} & \multicolumn{1}{c}{$\Omega_2$} &
\multicolumn{1}{c}{$\sqrt{\Omega_1^2 + \Omega_2^2}$} \\
\multicolumn{1}{c}{[mas/a]} & \multicolumn{1}{c}{[mas/a]} &
\multicolumn{1}{c}{[mas/a]} & \multicolumn{1}{c}{[mas/a]} &
\multicolumn{1}{c}{[mas/a]} \\
\noalign{\smallskip}
\hline
\noalign{\smallskip}
 \hphantom{-}0.0 &  7.19  &  \hphantom{-}1.26  & -2.99  &  3.24  \\
 -1.0 &  6.77  & \hphantom{-}0.59  & -2.14  &  2.22  \\
 -2.0 &  6.37  & -0.06  & -1.30  &  1.30  \\
 -3.0 &  5.96  & -0.72  & -0.46  &  0.85  \\
 -4.0 &  5.55  & -1.38  &  \hphantom{-}0.40  &  1.44  \\ \noalign{\smallskip}
 -2.9 &  5.98  & -0.68  & -0.50  &  0.84  \\
\noalign{\smallskip}
\hline
\end{tabular}
\end{minipage}
\begin{minipage}[b]{8.7cm}
\begin{center} cepheids \hphantom{OB stars} \end{center}
\vspace*{-2ex}
\begin{tabular}{ccccc}
\noalign{\smallskip}
\hline
\noalign{\smallskip}
\multicolumn{1}{c}{$\Delta e$} & \multicolumn{1}{c}{$\Omega_0$} &
\multicolumn{1}{c}{$\Omega_1$} & \multicolumn{1}{c}{$\Omega_2$} &
\multicolumn{1}{c}{$\sqrt{\Omega_1^2 + \Omega_2^2}$} \\
\multicolumn{1}{c}{[mas/a]} & \multicolumn{1}{c}{[mas/a]} &
\multicolumn{1}{c}{[mas/a]} & \multicolumn{1}{c}{[mas/a]} &
\multicolumn{1}{c}{[mas/a]} \\
\noalign{\smallskip}
\hline
\noalign{\smallskip}
 \hphantom{-}0.0 &  \hphantom{-}6.86  & -0.43  & -1.96  &  2.01  \\
 -1.0 &  6.43  & -0.93  & -1.21  &  1.53  \\
 -2.0 &  6.02  & -1.41  & -0.47  &  1.49  \\
 -3.0 &  5.59  & -1.90  &  \hphantom{-}0.29  &  1.92  \\
 -4.0 &  5.18  & -2.39  &  \hphantom{-}1.04  &  2.61  \\ \noalign{\smallskip}
 -1.5 &  6.23  & -1.17  &  \hphantom{-}0.84  &  1.44  \\
\noalign{\smallskip}
\hline
\end{tabular}
\end{minipage}
\end{center}
\end{flushleft}
\end{table*}

\begin{figure}
\leavevmode
\epsffile{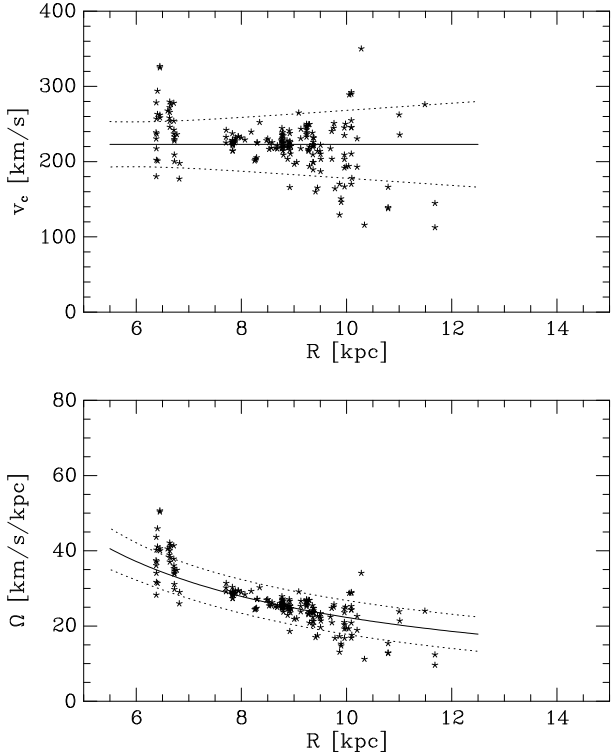}
\caption[]{\label{ob1} Rotation curve derived from the OB star data using the
full space velocities. The upper panel shows the circular velocity, whereas
the lower panel shows the angular velocity of the 197 OB stars associated
with 59 H\,{\sc ii} regions.
The solid lines indicate the rotation law of a flat rotation curve. The
dotted lines indicate error estimates based on the errors of the proper
motions, radial velocities and distances of the stars.}
\end{figure}

\begin{figure}
\leavevmode
\epsffile{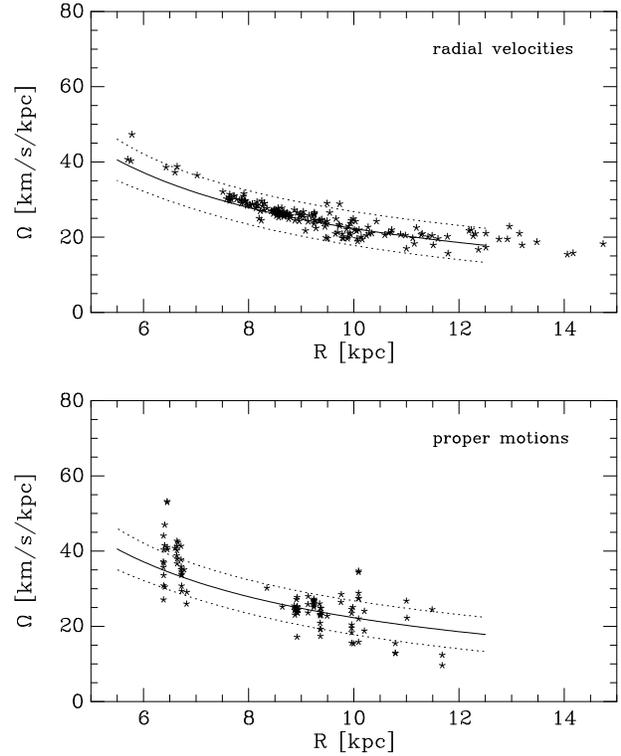}
\caption[]{\label{ob2}
Rotation curves derived from the OB star data using only radial or
tangential velocity components, respectively, assuming circular orbits of
the stars. There are 193 H\,{\sc ii} regions with measured radial
velocities (upper panel; this is essentially the rotation curve derived by
Brand \& Blitz (1993)\,) and 104 OB stars with known proper motions and
suitable projection angles.}
\end{figure}

\begin{figure}
\leavevmode
\epsffile{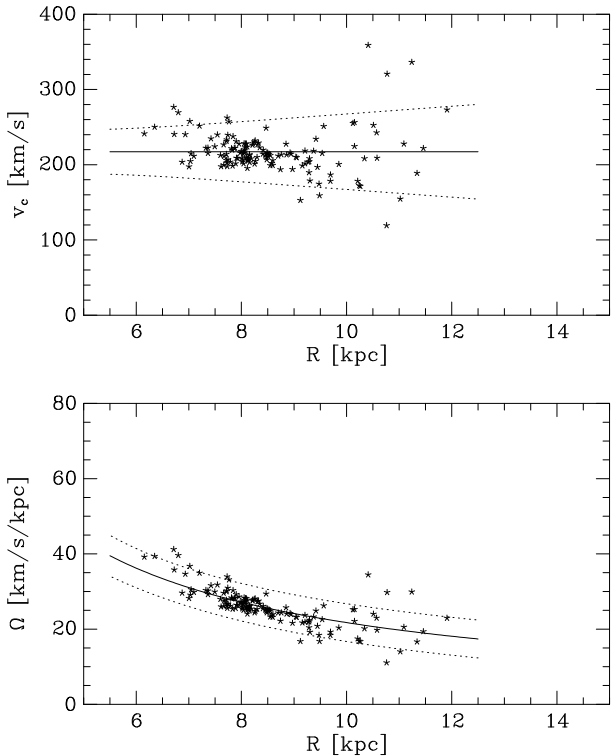}
\caption[]{\label{cep1}
Rotation curve derived from the cepheid data using the full space
velocities of 144 cepheids. The representation is the same as in
Fig.\,\protect\ref{ob1}.}
\end{figure}

\begin{figure}
\leavevmode
\epsffile{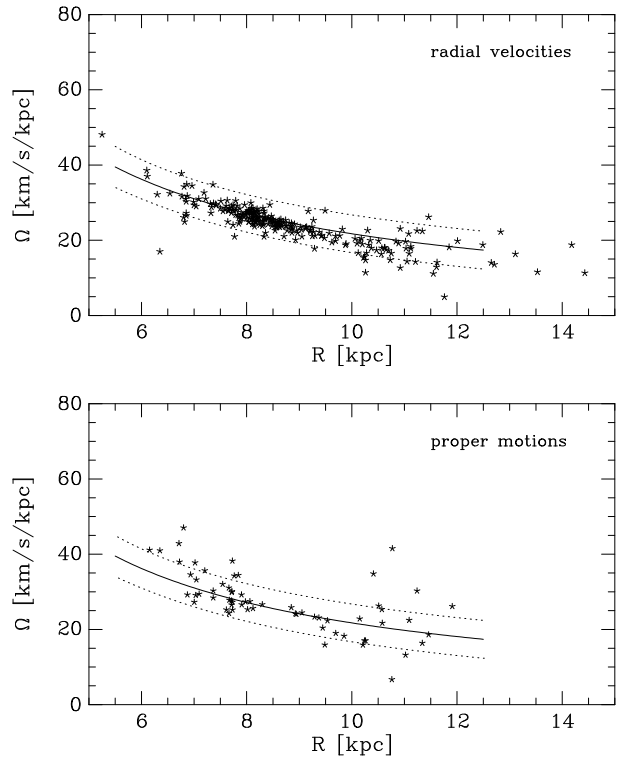}
\caption[]{\label{cep2} Rotation curve derived from cepheid data using only
radial velocities or proper motions, respectively. The upper panel shows
essentially the rotation curve from Pont et al.\ (1994) using radial
velocity components of 249 cepheids. The lower panel shows the rotation
curve using proper motions of 63 cepheids with suitable deprojection
angles.}
\end{figure}

\begin{figure}
\leavevmode
\epsffile{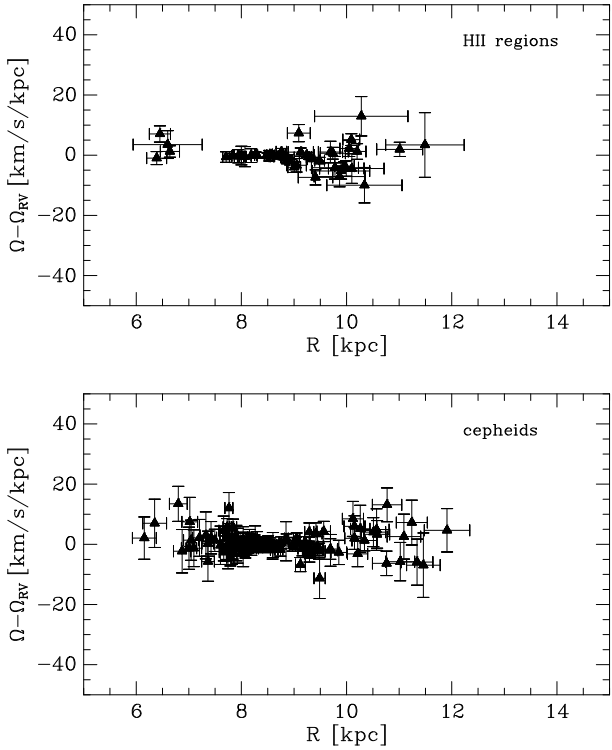}
\caption[]{\label{diff} The rotation curve derived solely from radial
velocities assuming circular orbits around the galactic centre, subtracted
{}from the rotation curve based on the space velocities of 48 H\,{\sc ii}
regions (upper panel, where stars belonging to the same H\,{\sc ii} region
have been grouped together) and 124 cepheids (lower panel). Stars in the
direction of the galactic centre or anticentre have been excluded.
Individual errors are indicated.}
\end{figure}

\begin{figure}
\begin{center}
\leavevmode
\epsffile{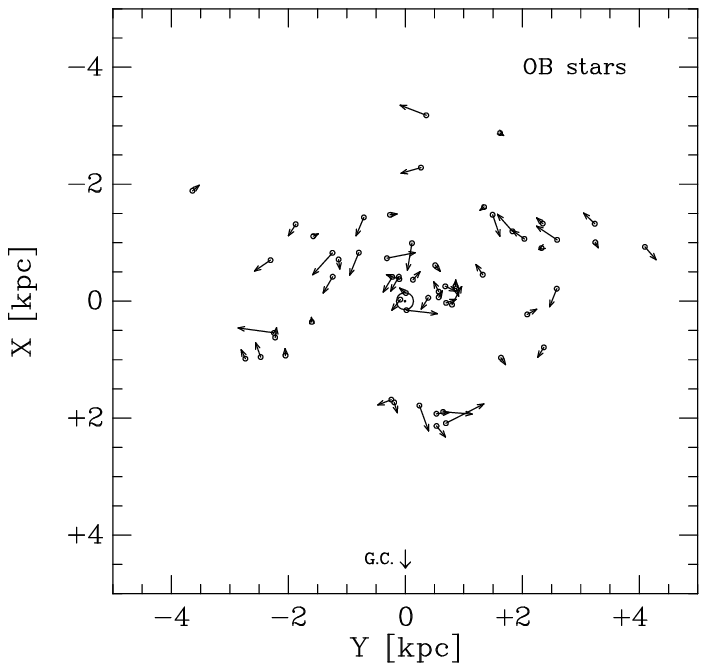}
\end{center}
\begin{center}
\leavevmode
\epsffile{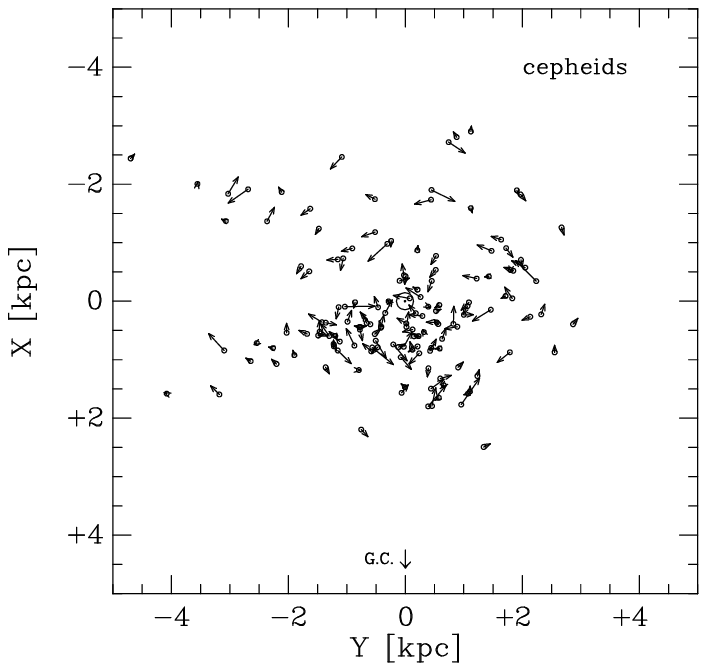}
\end{center}
\caption[]{\label{illu} Velocity residuals projected onto the galactic plane
after subtraction of a flat rotation curve, normalized by the individual
errors of the space velocities. A normalized residual of unit one
corresponds to a vector with a length of 150\,pc in the diagram. Notice the
aligned velocity pattern in the Perseus spiral arm in the upper right hand
quadrant for the OB stars (upper panel). The cepheids (lower panel) do not
show a similar feature.}
\end{figure}

\section{Data reduction}
\subsection{Angular velocity of the LSR}
In the FK5 system the velocities of the stars have a component of rigid
rotation due to the motion of the LSR around the galactic centre.
Assuming now that the stars move on circular orbits around the galactic
centre and that the rotation curve is flat, one may fit a model of the form
\begin{eqnarray} \label{model}
U & = & \Omega_0 R_0 \sin \varphi \nonumber\\
V & = & \Omega_0 R_0 ( \cos \varphi -1 )
\end{eqnarray}
to the space velocities of the stars. $U$ and $V$ correspond to the
directions of the $X$- and $Y$-axis of Fig.\,\ref{xyz} and denote the space
velocities of the stars which have been corrected for the solar motion with
respect to the LSR (cf.\ section \ref{solar motion}).
$R_0$ is the distance of the Sun from the galactic centre which we
assume as 8.5\,kpc throughout the present study, $\varphi$ denotes the
galactic azimuthal angle of a star.

Since some of the stars are fairly distant from the Sun we do not use
Oort's approximation but adopt rigorous formulae describing a flat rotation
curve. Deviations of the stars from the midplane are ignored.

{}From a $\chi^2$-fit of eq.\,(\ref{model}) to the space velocities of
the 197 OB stars we find
\begin{equation}
\Omega_0 = 5.5 \pm 0.4\,\mbox{mas/a}
\end{equation}
for the angular velocity of the LSR which corresponds to a circular
velocity of 223$\pm$14\,km/s in good agreement with the IAU 1985 value of
220\,km/s.

{}From the data of the 144 cepheids we obtain
\begin{equation}
\Omega_0 = 5.4 \pm 0.5\,\mbox{mas/a}
\end{equation}
in good agreement with the OB stars.

The estimated error of $\Omega_0$ has been calculated from the errors given
individually for the proper motions of each star in the PPM catalogue and
the errors of the radial velocities and the distances given in our
reference lists.
Furthermore we took into account the velocity dispersion of the stars
relative to the circular orbits by an additional error of 6.5\,km/s in both
velocity components for the OB stars (Brand \& Blitz 1993) and of 13\,km/s
in the U- and 9\,km/s in the V-direction for the cepheids (Pont et
al.\ 1994), respectively.

Judging from the numerical value of $\chi^2$ and the statistical
distribution of the velocity residuals we found that the error estimates
obtained this way were too low and have increased the errors of the proper
motions by 30\,\%. This is apparently due to systematic errors of the
proper motions in the PPM catalogue defined in the reference frame of the
FK5 system, which are not fully represented by the individual errors given
there.

\subsection{Influence of the reference system}
Moreover there is a further source of potentially serious errors of the proper
motions which is related to non-physical global rotations of the reference
system. There is now a general agreement that the IAU (1976) value of the
constant of precession, proposed by Fricke (1977) and used in the
construction of the
FK5, has to be corrected by about $\Delta p = -3.2 \pm 0.3$\,mas/a
(Williams et al.\ 1994). This corresponds to a rigid rotation with a spin
vector pointing to the ecliptical pole.

The immediate effects on our
results are illustrated in Table \ref{precession}. To each star the
correction has been applied and both data sets have been reduced again
according to the procedure described above. $\Omega_0$ is again the angular
velocity of the LSR wiht respect to the galactic centre. $\Omega_1$ and
$\Omega_2$ are angular velocities of the mean rotation of all stars about axes
lying in the galactic plane, pointing towards the galactic centre ($l=0\degr$)
and the direction of galactic rotation ($l=90\degr$), respectively.

As can be seen from Table \ref{precession} the introduction of the correction
of the constant of precession leads to an unacceptably large apparent
tilting of the galactic plane. Following Fricke (1977) we have therefore
considered a simultaneous correction of the motion of the vernal equinox
$\Delta e$. This corresponds to a rigid rotation of the reference frame about
an axis pointing towards the celestial pole. Table \ref{precession} shows that
the apparent tilting of the galactic plane can be minimized by adopting a
value of $\Delta e = -2.9$\,mas/a for the OB stars or of $\Delta e =
-1.5$\,mas/a for the cepheids. Nevertheless the remaining tilting is
significantly higher than in the
case where no correction at all was applied, i.\,e.\ 0.35\,mas/a for OB
stars with distances greater than 800\,pc and 1.10\,mas/a for cepheids.
In all these cases $\Omega_0$ agrees within 1\,mas/a with the values
given above. This discrepancy cannot be solved on the basis of the present
material.

The corrections $\Delta e$
of the motion of the equinox suggested here are of the same
order of magnitude as those proposed by Miyamoto and S\^oma (1993) and
by Wielen (unpublished). They do not agree exactly mainly
because of different samples of stars used. Miyamoto and S\^oma derived
$\Delta e = -1.2$\,mas/a from K giants. Wielen obtained $\Delta e =
-3.2$\,mas/a from 512 FK5 stars which were used by Fricke (1977) for
deriving the constant of precession and the motion of the equinox and which
were re-investigated by Schwan (1988) after the FK5 was completed.
We conclude that the spurious, non-physical rotations of the FK5 system seem
to be in total of the order of $\pm 1$\,mas/a in each component of the rotation
vector. This is consistent with the claimed accuracy of the FK5 system
with respect to non-physical rotations of about $\pm 0.7$\,mas/a (Fricke
1977, Schwan 1988).

Finally we note that the VLA measurement of the proper motion of the
Sgr\,A$^{*}$ radio source in the galactic centre of $-6.55 \pm 0.34$\,mas/a
(Backer 1996), implying after substraction of the peculiar velocity
of the Sun an angular velocity of the LSR of $\Omega_0 = 6.4$\,mas/a, deviates
only insignificantly from the values found here.

\subsection{Solar motion}
\label{solar motion}
Before applying Eq.\,(\ref{model}) the space velocities have to be corrected
for the solar motion with respect to the LSR. The standard values
for this motion are $U_0 = 9$\,km/s and $V_0 = 12$\,km/s (Delhaye 1965).

We introduced $U_0$ and $V_0$ as additional free parameters in
Eq.\,(\ref{model}) and checked the goodness of the fit for various
combinations of $U_0$ and $V_0$. The results are shown in Fig.\,\ref{grid}.

The $\chi^2$-fit to the OB star data rejects the standard values of 9 and
12\,km/s at a 4\,$\sigma$ confidence level, whereas the fit to the cepheid
data is in good agreement with the classical values. The minimum of
the numerical value of
$\chi^2$ for the cepheids however is quite flat and the errors are larger
than those for the OB stars, so we decided to use values of
$U_0 = 5$\,km/s and $V_0 = 6$\,km/s, which correspond within the errors
simultaneously to the minima of $\chi^2$ of both data sets.

Note that the value of $V_0$ is closely related to the asymmetric drift of
the sample of stars under consideration. Jahrei{\ss} and Wielen (1983) found
$V_0 = 5$\,km/s relative to the youngest stars in the solar neighbourhood
in close agreement with the value used here. The low
value of $U_0$ reflects the apparent inward motion of stars
relative to the LSR which was noted previously by Fich et al.\ (1989).
This effect is usually interpreted as an outward motion of the LSR
(Blitz \& Spergel 1991).

\subsection{Oort's constants}
We have determined Oort's constants $A$ and $B$ using stars within a circle of
1\,kpc around the Sun, assuming a constant radial gradient of
the rotation curve near the Sun. By means of a $\chi^2$-fit to the
circular velocity
\begin{equation}
v_{\mbox{\tiny c}}(R) = v_{\mbox{\tiny c}}(R_0) + \left( \frac{\mbox{d}
v_{\mbox{\tiny c}}}{\mbox{d} R} \right)_0 (R - R_0)
\end{equation}
we find for the OB stars $A=14.0\pm1.2$\,km/s/kpc and $B=-12.3\pm1.2
$\,km/s/kpc, which agree nicely with the flat shape of the rotation curve
over larger distance intervals. The cepheids give $A=15.8\pm1.6$\,km/s/kpc
and $B=-9.7\pm1.6$\,km/s/kpc. Obviously Oort's constants
only reflect the local behaviour of the rotation curve. Over larger distance
intervals the rotation curves show a flatter shape than one would assume
{}from the above values.

\section{Results and Discussion}
\subsection{Rotation curves}
Once the velocities have been corrected for the solar motion, it is
straightforward to determine the circular velocity $v_{\mbox{\tiny c}}(R)$ for
according to the formula
\begin{equation}
v_{\mbox{\tiny c}}=(U-\Omega_0 Y)\sin \varphi+(V+\Omega_0 X)\cos \varphi +
\Omega_0 R\,\,,
\end{equation}
where $R$ is the distance of the star from the galactic centre. It is now
no longer necessary to assume that the stars move on circular orbits
around the galactic centre, because the full space velocities are used.

Alternatively, assuming again circular orbits, one may derive rotation
curves based solely on radial or tangential velocity components of the
stars,
\begin{eqnarray} \label{rotrad}
\Omega_0(R) &=& \frac{v_{\mbox{\tiny rad}}}{R_0 \sin l \cos b} + \Omega_0
\qquad \\ \label{rotpm}
\Omega_0(R) &=& \frac{(\mu_l + \Omega_0) r \cos b}{R \cos (\varphi + l)} +
\Omega_0 \quad,
\end{eqnarray}
where $r$ denotes the distance of the star from the Sun,
$v_{\mbox{\tiny rad}}$ the radial velocity, and
$\mu_l$ the proper motion in the direction of galactic longitude $l$.
Both velocities have to be corrected again for the solar motion.
$r$ is the distance of the star from the Sun.

The resulting rotation curves determined using either the full space
velocities or deprojected radial or tangential velocity components are
shown in Figs.\,\ref{ob1} to \ref{cep2}. They agree very closely and are
all consistent with a flat shape of the rotation curve $\Omega(R)=\Omega_0
R_0 / R$.

Stars with unsuitable projection angles onto the supposed circular
velocities have been excluded. In the case where only radial velocity data are
used stars with $| \sin l\,| < 0.3$ have been rejected. If only proper
motion data are used stars with $| \cos(\varphi +l)\,| < 0.7$ have been
rejected. Therefore the overlap between these two samples is small and
we cannot correlate directly the rotation curve derived from radial velocities
with the rotation curve based on proper motions or both.

Instead, we show in Fig.\,\ref{diff} the deviations of circular velocities
derived from the space velocities from circular velocities based solely on
deprojected radial velocities of the stars illustrating again that both
methods give consistent results.

\subsection{Residual velocities}
The orientations of the velocity residuals after subtraction of the
systematic velocity components due to a flat rotation curve from the
space velocities of the stars are shown in Fig.\,\ref{illu} projected onto
the galactic plane. The velocity residuals are dominated by the errors of
the proper motions. Most of the velocity residuals can be shown to be
randomly orientated with the exception of the Perseus spiral arm. There is
a trend of coherent motion along the spiral arm with the effect that the OB
stars in the spiral arm tend to lag behind the general rotation of the
disk. Exactly such a behaviour is predicted by the density wave theory of
spiral structure for stars recently born in the shock front of the
interstellar gas included by the spiral density wave (Shu et al.\ 1972).
This is not observed for the cepheids. They are about 100 times older
than the OB stars and their systematic flow pattern is already dissolved
(Wielen 1979).

\end{document}